\documentclass{elsart}

\usepackage{graphicx}

\begin{document}

\begin{frontmatter}

\title{Nonlinear reaction with fractional dynamics}

\author{A.~A.~Stanislavsky\corauthref{cor1}}
\corauth[cor1]{E-mail: alexstan@ri.kharkov.ua}

\address{Institute of Radio Astronomy, 4 Chervonopraporna
St., Kharkov 61002, Ukraine}

\begin{abstract}
In this paper we consider a system of three fractional
differential equations describing a nonlinear reaction. Our
analysis includes both analytical technique and numerical
simulation. This allows us to control the efficiency of the
numerical method. We combine the numerical approximation of
fractional integral with finding zeros of a function of one
variable. The variable gives a desired solution. The solution has
a peak. Its position in time depends on the order of fractional
derivative. The two other solutions of this system have a
completely monotonic character.
\end{abstract}


\end{frontmatter}

\section{Introduction}
\thispagestyle{empty} The dynamical systems with fractional
dynamics are adequate for the description of various physical
phenomena in viscoelasticity, diffusion processes, dielectric
relaxation, etc \cite{1,2,3,4}. However, now the analysis of
nonlinear differential equations with fractional derivatives is in
early development. The point is that on the one hand the
derivation of such equations from some initial principles is in
the making. On the other hand the numerical simulation of
fractional differential equations are enough expensive~\cite{5}.
In the present state of affairs it may follow a formal way by
changing the derivatives of integer order in the well-known
nonlinear equations on the derivative of an arbitrary order. This
means a search to the touch together with overcoming the
difficulties of numerical analysis. So that was at the beginning
of fractional calculus. But there exists an alternative way. The
other approach may be based on the recent success in the study of
linear fractional differential equations. In particular, the
linear fractional equations have been well investigated
analytically in detail \cite{5a,5b,5c}. Moreover, the fractional
oscillator and the fractional relaxation have an clear physical
interpretation \cite{6,7}. Some nonlinear models turn out to be
very close to linear ones in a sense. This permits one to combine
analytical computations with simple numerical simulations. The aim
of this work is to present one of such nonlinear systems under
fractional dynamics.

The plan of the paper is as follows. Section~\ref{par2} is devoted
to a derivation of the nonlinear reaction obeying a stochastic
time arrow. The arrow leads to a system of fractional differential
equations that describe the process evolution. One of the
equations can be resolved by analytical methods. The sum of the
two others has also an clear analytical form. The feature analysis
is represented in Section~\ref{par3}. Nevertheless, the system of
the equations upon the whole is nonlinear. Therefore, the next
step is directed to numerical simulations of those equations what
cannot be resolved analytically. We rewrite them in the integral
form and apply a recursive procedure of numerical fractional
integration. It should be observed here that the results obtained
in Section~\ref{par3} proves to be useful for Section~\ref{par4}.
We sum up and discuss the outcome in Section~\ref{par5}.

\section{Theoretical model}\label{par2}
The nonlinear processes play an important role in the kinetics of
chemical reactions \cite{8}. The typical example is a reaction in
the gas that consists of atoms of two grades ($A$ and $B$). The
reaction $A+B=AB$ creates two-atom molecules. In the same time it
is not impossible an inverse process, dissociation $AB=A+B$. The
dynamical system can be described by differential equations in the
nonlinear form.

One of similar nonlinear reactions is mentioned in \cite{9}. Its
temporal evolution is written as
\begin{eqnarray}
\frac{d y_1}{dt}&=&-y_1\,,\nonumber\\ \frac{d
y_2}{dt}&=&y_1-y^2_2\,,\label{eq01}\\ \frac{d
y_3}{dt}&=&y^2_2\,,\nonumber
\end{eqnarray}
with the initial condition $N_1/N=y_1(0)=1$, $N_2/N=y_2(0)=0$,
$N_3/N=y_3(0)=0$, where $N$ is the common number of atoms, and
$N_1$, $N_2$, $N_3$ the number of molecules of three types arising
from the chemical reaction of the atoms. The equation for $y_1(t)$
is linear and is independent from $y_2(t)$ and $y_3(t)$. Using the
stochastic time arrow described in \cite{6}, the equation is
easily generalized in the fractional form.

The stochastic time arrow characterizes an interaction of a
physical system with environmental degrees of freedom. Then the
time variable becomes a sum of random temporal non-negative
intervals $T_i$. They are independent identically distributed
variables belonging to a L\'evy-stable distribution. Their sum
gives asymptotically a stable distribution with the index
$0<\alpha\leq 1$. Following the arguments of \cite{6,9a}, a new
time clock is defined as a continuous limit of the discrete
counting process $N_t=\max\{n\in{\bf N}: \sum_{i=1}^nT_i\leq t\}$,
where ${\bf N}$ is the set of natural numbers. The time clock is
the hitting time process $S(t)$. Its basic properties has been
represented in \cite{9a,9b}. The probability density of the
process $S(t)$ is written in the form
\begin{equation}
p^{S}(t,\tau)=\frac{1}{2\pi i}\int_{Br} e^{ut-\tau
u^\alpha}\,u^{\alpha-1}\,du\,, \label{eq01a}
\end{equation}
where $Br$ denotes the Bromwich path. This probability density has
a clear physical sense. It defines the probability to be at the
internal time $\tau$ on the real time $t$ (see details in
\cite{6}).

Let us make use of the general kinetic equation
\begin{equation}
\frac{d p(t)}{dt}=\hat{\bf W}p(t)\,,\label{eq02}
\end{equation}
where $p(t)$ defines the transition probability (or the
concentration of a substance) in a system, $\hat{\bf W}$ is the
transition operator between the system states. In a general case
the operator $\hat{\bf W}$ may be even nonlinear. Provided that it
is independent on time explicitly. Now we determine a new process
with the probability
\begin{equation}
p_\alpha(t)=\int_0^\infty
p^S(t,\tau)\,p(\tau)\,d\tau\,.\label{eq03}
\end{equation}
It is that the function $p^S(t,\tau)$ characterizes the stochastic
time arrow. The Laplace transform of the probabilities
$p_\alpha(t)$ and $p(t)$ gives the relation
$\tilde{p}_\alpha(s)=s^{\alpha-1}\tilde{p}(s^\alpha)$ with
$0<\alpha\leq 1$, and the Laplace image of $x(t)$ is written as
\begin{displaymath}
\tilde{x}(s)=\int_0^\infty e^{-st}x(t)\,dt\,.
\end{displaymath}
By the simple algebraic computations we have
\begin{displaymath}
\hat{\bf W}\tilde{p}_\alpha(s)=s^\alpha \tilde{p}_\alpha(s)-
p(0)s^{\alpha-1}\,.
\end{displaymath}
After the Laplace inverse transform the fractional representation
of Eq.(\ref{eq02}) takes the form
\begin{equation}
p_\alpha(t)=p(0)+\frac{1}{\Gamma(\alpha)}\int_0^t
d\tau\,(t-\tau)^{\alpha-1}\,
\hat{\bf W}p_\alpha(t)\,.\label{eq04}
\end{equation}
Although Eq.(\ref{eq04}) is integral, it can be expressed in terms
of fractional derivative \cite{6,9a,9b}. This approach permits one
to generalize Eqs.(\ref{eq01}) in that way too. For more details,
see the next section.

\section{Analytical calculations}\label{par3}
Consider the following nonlinear system described by fractional
differential equations \cite{10}
\begin{eqnarray}
\frac{d^\alpha y_1}{dt^\alpha}&=&-y_1\,,\nonumber\\ \frac{d^\alpha
y_2}{dt^\alpha}&=&y_1-y^2_2\,,\label{eq1}\\ \frac{d^\alpha
y_3}{dt^\alpha}&=&y^2_2\,,\nonumber
\end{eqnarray}
under the initial condition
\begin{displaymath}
y_1(0)=1,\quad y_2(0)=0,\quad y_3(0)=0,
\end{displaymath}
where $0<\alpha\leq 1$. Here the fractional operator
$d^\alpha/dt^\alpha$ is supposed in the sense of Caputo \cite{11}:
\begin{displaymath}
\frac{d^\alpha
x(t)}{dt^\alpha}=\frac{1}{\Gamma(n-\alpha)}\int^t_0\frac{x^{(n)}(\tau)}
{(t-\tau)^{\alpha+1-n}}\,d\tau, \quad n-1<\alpha<n,
\end{displaymath}
where $x^{(n)}(t)$ means the $n$-derivative of $x(t)$. To sum up
Eqs.(\ref{eq1}), we get
\begin{displaymath}
\frac{d^\alpha (y_1+y_2+y_3)}{dt^\alpha}=0\quad\Rightarrow\quad
y_1+y_2+y_3=1\,.
\end{displaymath}
The system analysis shows above all that the first equation for
$y_1$ is linear. It may be integrated analytically, namely
$y_1(t)=E_\alpha(-t^\alpha)$. The variable $y_1$ is expressed in
term of the one-parameter Mittag-Leffler function $E_\alpha(z)$.
Thus, the value $y_1$ becomes vanishingly small with $t\to\infty$
as $t^{-\,\alpha}/\Gamma(1-\alpha)$. On the other hand, from the
second and third equations it follows that the sum of $y_2$ and
$y_3$ gives
\begin{displaymath}
\frac{d^\alpha (y_2+y_3)}{dt^\alpha}=y_1\,.
\end{displaymath}
This means that the definition of $y_1$ allows one to establish
the functional dependence between $y_2$ and $y_3$. Therefore, the
sum takes the form
\begin{displaymath}
y_2+y_3=1-E_\alpha(-t^\alpha)\,.
\end{displaymath}
This result tends to one as $t\to\infty$.

We find it useful to define the identity
\begin{displaymath}
E_{\alpha,1}(-t^\alpha)+t^\alpha
E_{\alpha,\,1+\alpha}(-t^\alpha)=1\,,
\end{displaymath}
where $E_{\alpha,\,\beta}(z)=\sum_{n=0}^\infty z^n/\Gamma(\alpha
n+\beta)$ is the two-parameter Mittag-Leffler function, and
$E_\alpha(z)$ is a particular case of $E_{\alpha,\,\beta}(z)$,
namely $E_{\alpha,1}(z)$. The Laplace image of $t^\alpha
E_{\alpha,1+\alpha}(-t^\alpha)$ is written as
\begin{displaymath}
\int_0^\infty e^{-st}\,t^\alpha
E_{\alpha,1+\alpha}(-t^\alpha)\,dt=\frac{1}{s}\,\frac{1}{(1+s^\alpha)}\,,
\end{displaymath}
while
\begin{displaymath}
\int_0^\infty e^{-st}\,
E_{\alpha,\,1}(-t^\alpha)\,dt=\frac{s^{\alpha-1}}{(1+s^\alpha)}\,.
\end{displaymath}
The sum of these images reduces to $1/s$.

In the upshot it still remains to find the variable $y_2$. The
value satisfies a nonlinear fractional differential equation with
the right-hand term. In other words this equation is
inhomogeneous:
\begin{equation}
\frac{d^\alpha
y_2}{dt^\alpha}+y^2_2=E_\alpha(-t^\alpha)\,.\label{eq2}
\end{equation}
Its form is such that for the final conclusion it is necessary to
provide a numerical simulation. Nevertheless, our analytical
analysis will not be full without asymptotic estimations as
applied to $y_2$. If $\alpha$ is equal to one, we can neglect the
right-hand term because of its exponential decay. The value
$y_2(t)$ behaves algebraic for $t\gg 1$. If the index $\alpha$ is
otherwise ($0<\alpha<1$), the right-hand term of Eq.(\ref{eq2}) is
the one-parameter Mittag-Leffler function having an algebraic
asymptotic form. What term in Eq.(\ref{eq2}) gives a significant
contribution in a given case? The feature will be established by a
numerical simulation more clearly. The variable $y_2(t)$ becomes
vanishingly small, when $t$ tends to infinity.

\begin{figure}
\centering
\includegraphics[width=12 cm]{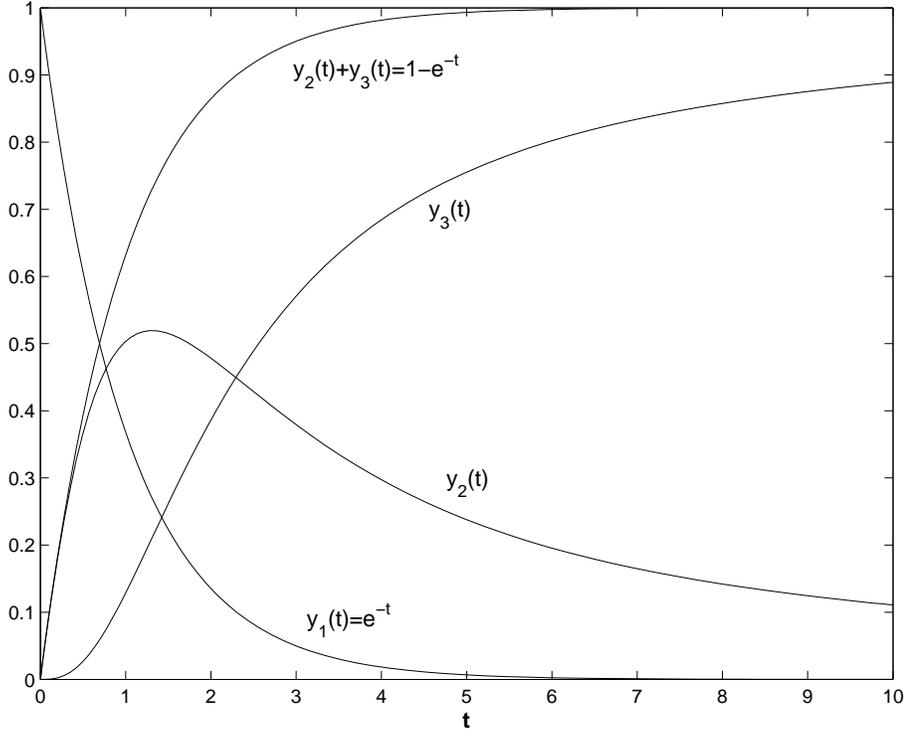}
\caption{\label{fig1}Nonlinear reaction with ordinary time
evolution (\ref{eq01})\,.}
\end{figure}

Moreover, the Adomain's composition method assumes a series
solution for $y_2$ \cite{10}. The first three terms of this series
are given by
\begin{equation}
y_2(t)=\frac{t^\alpha}{\Gamma(1+\alpha)}-\frac{t^{2\alpha}}
{\Gamma(1+2\alpha)}+\left(1-\frac{\Gamma(2\alpha+1)}
{\Gamma^2(\alpha+1)}\right)\frac{t^{3\alpha}}{\Gamma(3\alpha+1)}
\,\dots\,.\label{eq4}
\end{equation}
They indicate the presence of a maximum. However, the
determination of its properties (position and value) is not enough
reliable for the composition method. The point is that the Taylor
series converges only for $0<t<1$, but the maximum is the case for
$t>1$. Thus the estimation (\ref{eq4}) for the maximum will not be
exact. This can be easy to verify numerically for $\alpha=1$,
using the usual methods (for example, the Runge-Kutta technique or
something like that). In this case, the system (\ref{eq1})
consists of first-order differential equations. Their numerical
solutions are represented in Fig.~\ref{fig1}\,.

\section{Numerical approximation}\label{par4}

To solve the nonlinear fractional differential equation, the
principal approach is to use a numerical approximation of
fractional integral. Its advantage is due to the fact that the
kernel of this integral is regular for $\alpha>1$, and weekly
singular for $0<\alpha<1$, hence integrable (at least in the
improper sense) for any index $\alpha>0$. In contrast, the kernel
of fractional derivative is not integrable, and it requires
special regularization methods restricting possibilities of the
numerical simulation \cite{12}.

Let $\{t_n=nh: n=0,1,2,\dots,N\}$ be a partition of the interval
$[0\ \ T]$ into a grid, where $h=T/N$. According to \cite{12}, the
fractional integral of the value $y$ is computed as
\begin{equation}
J^\alpha y_N=\frac{h^\alpha}{\Gamma(2+\alpha)}\sum_{n=0}^N
c_{n}^{(N)}y_n\,, \label{eq5}
\end{equation}
where the quadrature weights are derived from a product
trapezoidal rule to the following:
\begin{displaymath}
c_{n}^{(N)}=\cases{(1+\alpha)N^\alpha-
N^{\alpha+1}+(N-1)^{\alpha+1}\,,&if $n=0$;\cr 1\,,&if $n=N$;\cr
(N-n+1)^{\alpha+1}-2(N-n)^{\alpha+1}+ & \cr
+(N-n-1)^{\alpha+1}\,,& if $0<n<N-1$.\cr}
\end{displaymath}
The error of this algorithm is of order $O(h^2)$. The accuracy can
be improved by the Richardson extrapolation procedure \cite{12}.

\begin{figure}
\centering
\includegraphics[width=12 cm]{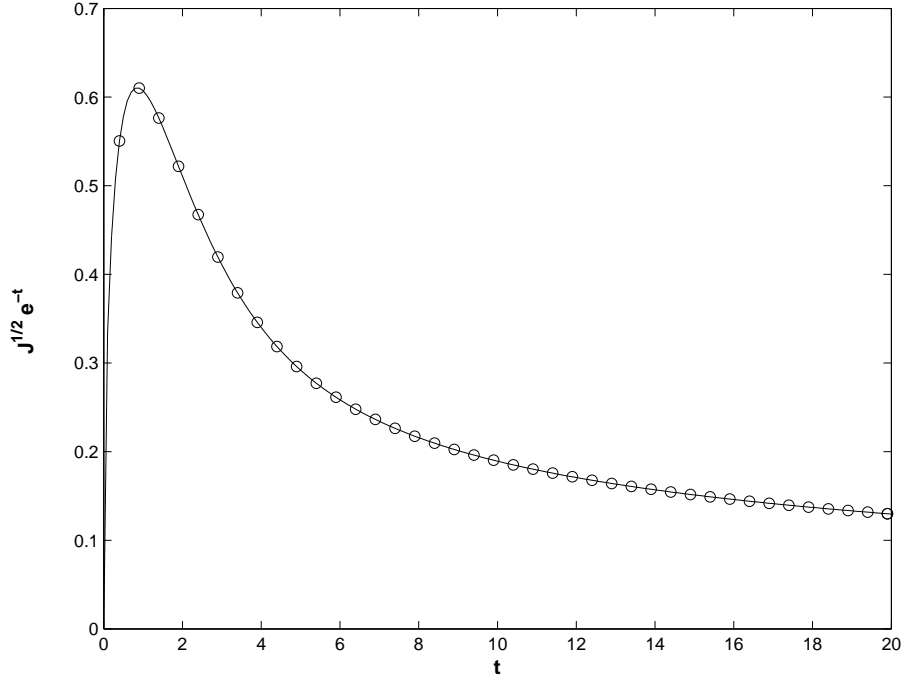}
\caption{\label{fig2}Fractional one-half-order integral of the
exponential function, $J^{1/2} \exp(-t)$, obtained from
Eq.(\ref{eq6}) (line) and by means of Eq.(\ref{eq5}) (circles).}
\end{figure}

As an example to this algorithm, we suggest to consider the
fractional one-half-order integral of the exponential function.
The integral has an analytical representation:
\begin{eqnarray}
J^{1/2}\exp(-t)&=&\frac{1}{\Gamma(1/2)}\int_0^t
(t-\tau)^{-1/2}\,e^{-\,\tau}\,d\tau=\nonumber\\
&=&\frac{1}{\sqrt{\pi t}}\int_0^\infty
e^{-\,\tau^2/(4t)}\,\sin\tau\,d\tau=e^{-\,t}
\mathrm{erfi}(\sqrt{t})\,,\label{eq6}
\end{eqnarray}
where $\mathrm{erfi}(x)=\frac{2}{\sqrt{\pi}}\int_0^x\exp(y^2)\,dy$
is the imaginary error function. The transcendental function is
expanded into a Taylor series about $x$:
\begin{displaymath}
\mathrm{erfi}(x)=\frac{2}{\sqrt{\pi}}\sum_{m=0}^\infty
\frac{x^{2m+1}}{(2m+1)m!}\,.
\end{displaymath}
The expansion allows one to compare $J^{1/2}\exp(-t)$ in the
analytical form (\ref{eq6}) with the numerical simulation result
of the fractional integral by means of the above-mentioned
algorithm (\ref{eq5}) (see Fig.~\ref{fig2}).

\begin{figure}
\centering
\includegraphics[width=12 cm]{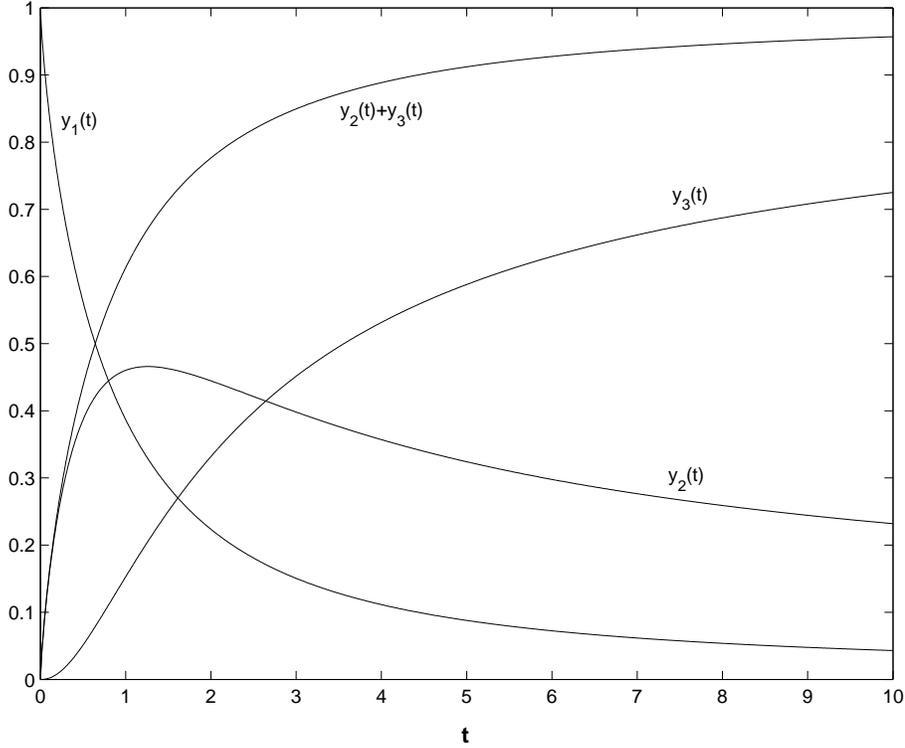}
\caption{\label{fig3}Nonlinear reaction with time evolution under
fractional derivative with the index $\alpha=0.8$\,.}
\end{figure}

Next we transform Eq.(\ref{eq2}) into the integral form:
\begin{equation}
J^\alpha\frac{d^\alpha y_2(t)}{dt^\alpha}+J^\alpha
y^2_2(t)=y_2(0)+y_2(t)+J^\alpha
y^2_2(t)=1-E_\alpha(-t^\alpha)\,.\label{eq7}
\end{equation}
It is useful to remark the following property: $0\leq
y_2(t)+J^\alpha y^2_2(t)\leq 1$. As $J^\alpha y^2_2(t)\geq 0$, it
is easy to find $0\leq y_2(t)\leq 1$. The upper limit of $y_2(t)$
is too strong. As will be shown below, $y_2(t)$ is less than one.
The positivity property of $y_2(t)$ at any time is very important.
The same exists also for $y_1(t)$ and $y_3(t)$. Both functions
$y_1(t)$ and $y_3(t)$ are completely monotonic, but $y_2(t)$ is
not. The latter has a maximum.

Applying the approximation (\ref{eq5}), Eq.(\ref{eq7}) is written
as
\begin{displaymath}
y_2(Nh)+\frac{h^\alpha}{\Gamma(2+\alpha)}\sum_{n=0}^N
c_{n}^{(N)}y_2^2(nh)=1-E_\alpha(-N^\alpha h^\alpha)\,.
\end{displaymath}
The expression takes the form of an ordinary quadratic equation:
\begin{eqnarray}
\frac{h^\alpha}{\Gamma(2+\alpha)}\,y^2_2(Nh)&+&y_2(Nh)+\frac{h^\alpha}
{\Gamma(2+\alpha)}\sum_{n=0}^{N-1}
c_{n}^{(N)}y^2_2(nh)-\nonumber\\ &-&1+E_\alpha(-N^\alpha
h^\alpha)=0\,.\label{eq8}
\end{eqnarray}
Then the recursive procedure is obtained in an elementary manner.
The numerical scheme permits one to calculate $y_2(nh)$ from
$y_2((n-1)h)$ to solve the quadratic equation. Each subsequent
value of $y_2$ depends on previous ones, because the fractional
integration exhibits a long-term memory loss. For computing the
one-parameter Mittag-Leffler function $E_\alpha(-t^\alpha)$ we use
a numerical scheme listed in Algorithm 4 \cite{12} that is
reproduced with minor corrections from \cite{12a}. The quadratic
equation (\ref{eq8}) has two solutions. One of them is positive,
and another is negative. We select the non-negative solution in
the force of the above analysis to the analytical properties of
$y_2(t)$. The temporal evolution of this system for $\alpha=0.8$
is shown in Fig.~\ref{fig3}. Formally, the system behavior for
$0<\alpha<1$ is like to the system one for $\alpha=1$: the
variable $y_1(t)$ monotonically tends to zero, the variable
$y_3(t)$ to one, and the variable $y_2(t)$ possesses one maximum
after which it decays to zero. But there are differences that we
discuss in the next section.

\begin{figure}
\centering
\includegraphics[width=12 cm]{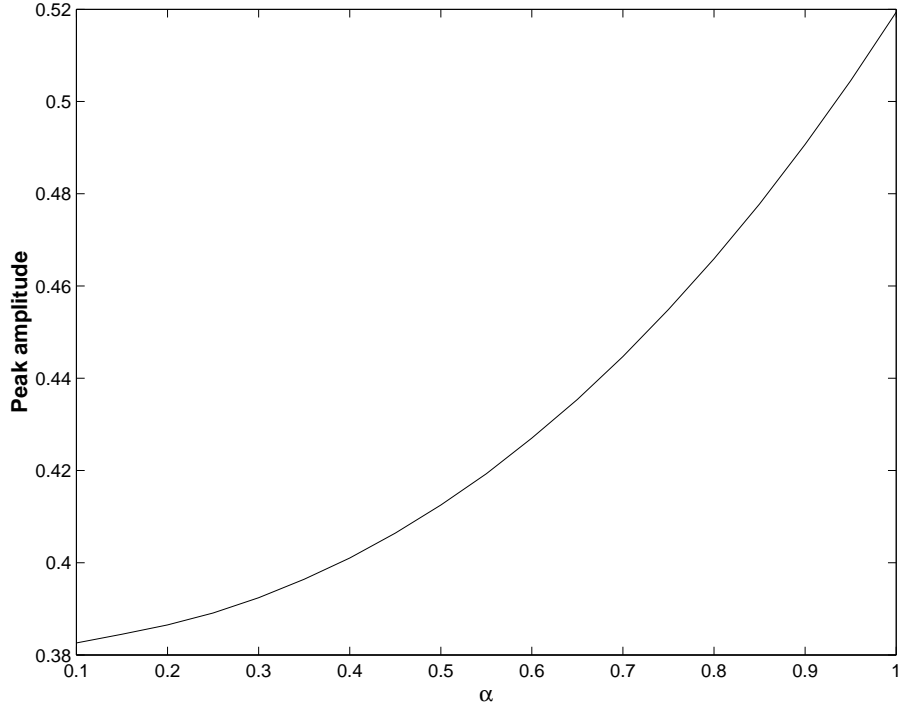}
\caption{\label{fig4}Peak amplitude of the variable $y_2$ for
$0.1\leq\alpha\leq 1$ in the nonlinear reaction with fractional
dynamics\,.}
\end{figure}

This approach to the numerical simulation of the system
(\ref{eq1}) has the following advantages. Firstly, the numerical
errors are of the order $O(h^2)$. If we use Algorithm 3 from
\cite{12}, they would be of the order $O(h^{1+\alpha})$, i.\ e.\
they would be slightly more. Moreover, Algorithm 3 is not less
expensive than the procedure (\ref{eq5}). A reduction in the
amount of computational work can be achieved by using a graded
mesh, thereby taking the O($N^2$) method of (\ref{eq5}) and making
it O($N\log N$). This will be the same as in Algorithm 3. Thirdly,
the Richardson extrapolation procedure, improving the numerical
simulation accuracy, does not have been proved still for Algorithm
3. Hence, its application to Algorithm 3 is only based on
assumptions of its expected correctness \cite{12}. As for the
procedure (\ref{eq5}), the use of the Richardson extrapolation is
exactly true for any $\alpha>0$. It is also worth noticing that
Algorithm 3 itself cannot yield a numerical solution of nonlinear
fractional equations. The algorithm should be added by calculating
zeros of the corresponding function of one variable. It is that
the variable is a required solution. However, this way to
numerical simulations of nonlinear fractional equations is not
considered in the review \cite{12}. In our opinion the approach
has good perspectives.

\section{Results and Discussion}\label{par5}

The variables $y_1$, $y_2$ and $y_3$ describe  a relaxation of
substances in some physical system more than anything yet. And
what is why. One of its possible interpretations lies in the
representation of the process as a reaction. Some substance
promotes an initiation of two other interacting substances. One of
them is metastable and transforms into another. But this reaction
does not occur at once. It evolutes in time. Therefore the change
of the variables $y_1$, $y_2$ and $y_3$ manifests something like a
relaxation.

\begin{figure}
\centering
\includegraphics[width=12 cm]{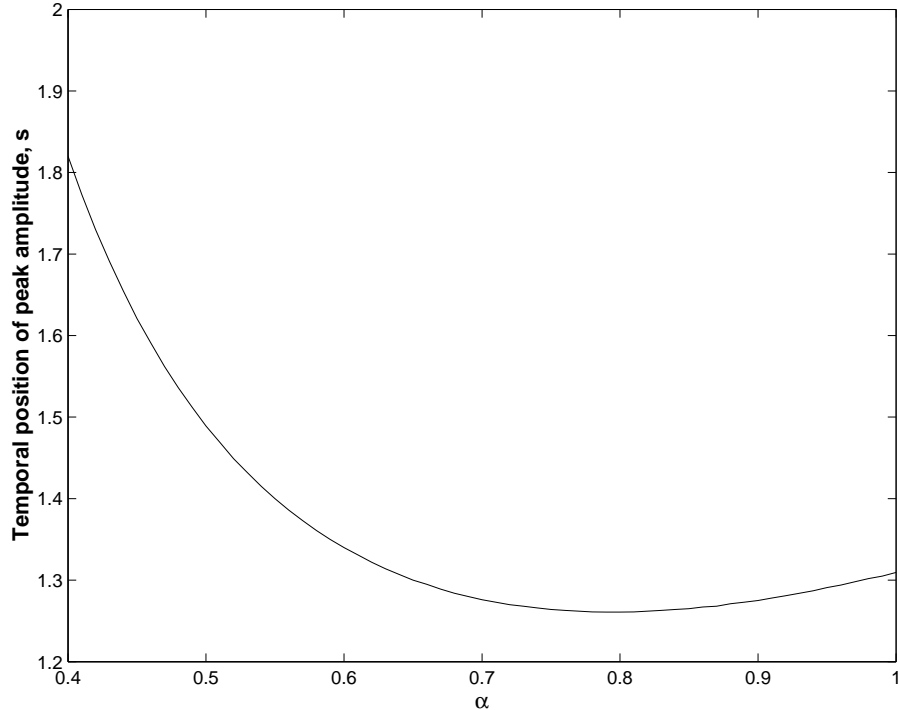}
\caption{\label{fig5}Time position of the $y_2$'s maximum in the
dependence of the index $\alpha$ for the nonlinear reaction with
fractional dynamics\,.}
\end{figure}

Let us choose the index $\alpha$ as a governing parameter. This
means that we will change it from zero to one for establishing how
it influences on the character of the system (\ref{eq1}). The key
characteristics of this system is a temporal position of the
$y_2$'s maximum and its value. The results can be only obtained by
numerical simulations. They are represented in Fig.~\ref{fig4} and
Fig.~\ref{fig5}. The first observation shows that the model
(\ref{eq1}) with the index $\alpha=1$ is well agreed with the
models for $\alpha\neq 1$. Otherwise speaking, all they together
form a smooth line in the interval $0<\alpha\leq 1$.
Figure~\ref{fig5} has been shown for $0.4\leq\alpha\leq 1$ to be
easy to recognize a minimum in this dependence on $\alpha$.
Really, the minimum is the case for $\alpha\approx 0.8$. This
indicates that at first the temporal position comes to the
abscissa axis, when the index $\alpha$ changes from one to $\sim
0.8$. However, the next decay evolution of the index leads to the
motion of the $y_2$'s maximum outside of the abscissa axis. The
peak amplitude has a completely monotonic character depending on
$\alpha$ (see Fig.~\ref{fig4}).

By the concrete example we have shown that the numerical procedure
of fractional integration can provide a convergent method for the
solution of nonlinear systems of fractional differential
equations. The approximation series solution (Adomian
decomposition) is not so successful because of the too slow
convergence of the corresponding series expansion for such a type
of equations. Therefore, in this case the decomposition plays only
a secondary role. Although the approximation series solution can
be obtained to any desirable number of terms, this makes the
method expensive and ineffective. Probably, it would be better to
add the algorithm by an asymptotic expansion. However, it is
difficult to suggest any universal method. Thus, the numerical
simulations turn out to be especially useful for the analysis of
nonlinear fractional differential equations.

\section*{Acknownledgement}
The author expresses his sincere appreciation to Prof. K. Diethelm
and Prof. N.J. Ford for their help.

\end{document}